\documentclass[ ]{copernicus2}  

\usepackage{color}
\usepackage[T1]{fontenc}

\begin{document}

\title{Classical Higgs mechanism in plasma
}

\author[1]{R. A. Treumann}
\author[2]{W. Baumjohann}

\affil[1]{International Space Science Institute, Bern, Switzerland}
\affil[ ]{\centerline{$^2$Space Research Institute, Austrian Academy of Sciences, Graz, Austria}\\
\emph{Correspondence to}: Wolfgang.Baumjohann@oeaw.ac.at}

\runningtitle{Plasma waves and Higgs fields}

\runningauthor{R. A. Treumann, W. Baumjohann}

\received{ }
\revised{ }
\accepted{ }
\published{ }


\firstpage{1}

\maketitle

  

\noindent\emph{Abstract}.-- It is demonstrated that in high temperature collisionless plasmas the propagation of high-frequency electromagnetic waves is naturally subject to a classical Higgs mechanism which in some cases generates a very small though finite mass on the otherwise massless photons. Though this has no further obvious consequence for any plasma processes, it is very satisfactory because it explains the retardation and cut-off of electromagnetic waves when encountering plasmas. Interestingly, low frequency electromagnetic plasma waves (Alfv\'en waves and whistlers) remain carried by confined massless photons. These just experience the increase in the dielectric constant provided by the plasma in the low frequency range below the cyclotron frequency.

\section{Introduction}
For anyone who is not firm in the mathematics of Abelian and non-Abelian field theory the Higgs (more appropriately Anderson-Brout-Englert-Guralkin-Hagen-Higgs-Kibble-Mi\-g\-d\-al-Nambu-Polyakov) mechanism (to quote all of its quite independent inventors in alphabetical order), by which a massless particle gains a finite mass \citep{anderson1963,englert1964,guralnik1964,higgs1964,migdal1967,nambu1960}, looks fairly mysterious. However, the basic idea is, in principle, quite simple. \citep[Its origin goes back to][his formulation of the equation of massive bosons, and  the short range meson theory of Yukawa.]{proca1936} Any massless particle, when interacting with a large-scale background field in some viscous way will become retarded and gain inertia, i.e. mass. \citep[The coupling factors $q_\mathit{yuk}$ of the two fields are called after][who introduced them first in 1935.]{yukawa1949} In fact one may wonder whether this mechanism also works in classical field theory. \citet{schwinger1962} and \citet{anderson1963}  already pointed on some similarity between quantum field theory and plasmons. In the following we show that the Higgs mechanism indeed works sometimes in a plasma to provide mass to otherwise massless photons.

\section{HF photon-plasma linear interaction}  
Consider the interaction between an electromagnetic wave and a collisionless unmagnetised dilute plasma of some density $N$. The dispersion relation of such transverse waves in a plasma \citep[found in any textbook on plasma waves, e.g.][chpt. 9]{baumjohann1996} is very simply obtained by straightforward calculation:
\begin{equation}
\mathcal{N}^2\equiv\frac{k^2c^2}{\omega^2}=1-\frac{\omega_e^2}{\omega^2}
\end{equation}
Here $\mathcal{N}$ is the refraction index, $k$ wavenumber, $\omega$ frequency, and $\omega_e=\sqrt{e^2N/\epsilon_0m_e}$ the electron plasma frequency at number density $N$. Nothing special seems to be about this expression which, when resolving it for the frequency, produces
\begin{equation}
\omega^2=k^2c^2+\omega_e^2
\end{equation}
In this version it shows that electromagnetic waves in a plasma for plasma frequency small against the wave frequency are free space modes propagating at light velocity $c$, while at lower frequency they  become retarded. Transforming this dispersion equation into an operator equation one easily obtains the wave equation for the electromagnetic field, say the vector potential $\delta\vec{A}$, as
\begin{equation}
\bigg(\nabla^2-\frac{1}{c^2}\frac{\partial^2}{\partial t^2}\bigg)\:\delta\vec{A} =\frac{\omega_e^2}{c^2}\:\delta\vec{A}
\end{equation}
This is the usual electrodynamic wave equation with a symmetry breaking term on the right which depends on the properties of the plasma and couples the plasma to the vector potential $\delta\vec{A}$. 

We are speaking here about waves; for this reason we use the vector potential in its fluctuation form $\delta\vec{A}$ as a wave field obeying the wave equation. Clearly, it is the term on the right-hand side of this equation which causes the retardation of the wave and the deviation of its speed from $c$ as a function of frequency $\omega$ respectively wavenumber $k$. The symmetry breaking term thus should give rise to a small mass $M_\mathit{ph}$ of the photons. This is indeed seen, when comparing it with Proca's equation for heavy bosons. Then the factor of the wave field on the right should be equal to some yukawa-coupling constant $q_\mathit{yuk}$ times the so far unknown background field. This term gives the ratio of the photon momentum to the quantum of action
\begin{equation}
\frac{\omega_e}{c}= \frac{M_\mathit{ph}c}{\hbar}\quad\mathrm{or}\quad M_\mathit{ph}=\frac{\hbar\omega_e}{c^2}
\end{equation}
Clearly the photon mass induced by the presence of the plasma is rather small. It is of the order of the energy contained in one Langmuir plasmon $\hbar\omega_e$ and, for high wave frequencies, does not count at all. It comes into play only at wave frequencies $\omega\gtrsim\omega_e$ close to the plasma frequency where the photons due to the gain in mass are strongly retarded, assuming long wavelengths and finally cannot penetrate deeper into the plasma. This brings up the question for the field which is responsible for the symmetry breaking term, and what is the coupling coefficient which couples the fluctuating wave field to the massive plasma field. Both these questions can, in the case of the plasma, quite easily be answered.

To this end we need to find the wave equation of the massive plasma field (which by the Higgs assumption is not massless but, in contrast to the gauge field, carries a mass). This means, we must derive the above wave equation under the assumption of an interaction between the plasma and the massless electromagnetic field. This is done when realising that the interaction between the plasma and the photon field implies the presence of an induction current. The wave equation then assumes the general form
\begin{equation}
\bigg(\nabla^2-\frac{1}{c^2}\frac{\partial^2}{\partial t^2}\bigg)\:\delta\vec{A} = -\mu_0\delta\vec{J} 
\end{equation}
where $\delta\vec{J}=-e\delta(N\vec{u})$ is the induced current. It is carried by electrons of charge $-e$ moving with bulk velocity $\vec{u}$ against the ions under the action of the wave potential field $\delta\vec{A}$. This equation can formally be written 
\begin{equation}
\bigg(\nabla^2-\frac{1}{c^2}\frac{\partial^2}{\partial t^2}\bigg)\:\delta\vec{A} = q_\mathit{yuk}H_0\delta\vec{A} 
\end{equation}
when introducing the Yukawa coupling constant. Here $H_0$ is the large  homogeneous and constant non-fluctuating value of the classical Higgs field which is coupled to the fluctuations of the gauge field $\delta\vec{A}$ through $q_\mathit{yuk}$.

What concerns the plasma dynamics, it suffices to consider the electron fluid equations in their linearised form which is the only version which leads to a wave equation for the massive electron fluid, assuming quasi-neutrality. These are the continuity, electron momentum conservation, and Poisson's equation taken in the radiation gauge. Expanding all quantities into background and fluctuations, we have $N=N_0+\delta N, \vec{u}=\vec{u}_0+\delta\vec{u}, \vec{u}=0$. We allow for a finite temperature of the electrons $T_e=m_ev_e^2/2$, with electron thermal velocity $v_e$, which leads to a finite pressure gradient in the electron momentum equation. As is well known, all linearised equations can be reduced to a wave equation, for instance for the density fluctuation, given as
\begin{equation}
\bigg(\nabla^2-\frac{1}{v_e^2}\frac{\partial^2}{\partial t^2}\bigg)\:\delta N = \frac{\omega_e^2}{v_e^2}\delta N~~\mathrm{where}~~\omega_e^2=\frac{e^2N_0}{\epsilon_0m_e}
\end{equation}
is the plasma frequency with respect to the background density $N_0$. This equation is the wave field equation of the massive field fluctuations which interact with the massless photon field, just what is required in the classical Abelian theory of the Higgs mechanism. The coupling, on the other hand, is provided by the current term in the photon field equation. It is, however, seen already that the average background field which will provide the mass term in the otherwise massless field equation, is in this case the average density field $H_0=N_0$. Since this is large, while all other fluctuation fields $\delta N, \delta\vec{u}, \delta\vec{A}$ are small fluctuating quantities, all second order terms in the fluctuations can be neglected. Then only terms containing linear fluctuations and the zero-order massive field term $N_0$ will be retained and survive. 

Following this procedure in the calculation of the current fluctuation in the photon field wave equation, we have $\partial_t\delta\vec{J}=-eN_0\partial_t\delta\vec{u}$, and we can use the electron momentum to express the time derivative of the velocity fluctuation (for generality including the adiabatic index $\gamma$)
\begin{equation}
\partial_t\delta\vec{u}=-\frac{e}{m_e}\delta\vec{E}-\frac{\gamma T_e}{m_eN_0}\nabla\delta N
\end{equation}
Combining with Poissons equation $\nabla\cdot\delta\vec{E}=-(e/\epsilon_0)\delta N$ and taking the time derivative of the wave field equation in order to produce the time derivative of the current on its right, we find
\begin{equation}
\mu_0\,\partial_t\delta\vec{J}= -\bigg[\frac{e^2N_0}{m_e}\mu_0+\frac{\gamma T_e}{m_ec^2}\nabla^2\bigg]\:\partial_t\delta\vec{A}
\end{equation}
Finally, inserting this into the photonic wave field equation and rearranging yields the wanted result which even contains a small thermal correction term which usually plays no role:
\begin{equation}
\bigg[\nabla^2-\frac{1}{c^2}\bigg(1-\frac{\gamma v_e^2}{c^2}\bigg)^{-1}\bigg]\delta\vec{A}=\frac{\omega_e^2}{c^2}\bigg(1-\frac{\gamma v_e^2}{c^2}\bigg)^{-1}\delta\vec{A}
\end{equation}
The quantity which contains the large stationary massive background field, which in this case the photons have interacted with, is the density field $N_0$ of the plasma itself. By the massive wave equation, the density is subject to fluctuations.  However, the coupling with $N_0$ is the crucial action of the massive field because like in the quantum Higgs mechanism it is the agent which generates the mass on the photons. 

What remains is the determination of the yukawa coupling coefficient for the two fields $N_0$, the large and on the fluctuation scale homogeneous background field, and the electromagnetic fluctuation field $\delta\vec{A}$. This is simple matter as it can be read directly from the right-hand side of the last expression. When neglecting the thermal correction term which anyway is very close to unity, we find
\begin{equation}
q_\mathit{yuk}= \frac{e^2}{\epsilon_0m_ec^2}=\frac{e^2\mu_0}{m_e}= \frac{a_0}{\epsilon_0}\approx 3\times10^{-14}~~\mathrm{m}
\end{equation}
Here $a_0$ is the classical electron radius. This term contains only natural constants and thus is a constant itself. However, by coupling to the photon field, together with the plasma response, which is contained in the plasma frequency, it gives rise to the small but finite photon mass 
\begin{equation}
M_\mathit{ph}\approx 7\times 10^{-47} \frac{N_0\,[\mathrm{m}^{-3}]}{10^6} ~~\mathrm{kg}
\end{equation}
which is otherwise completely determined by the background density field $N_0$ which enters the plasma frequency. This mass gives rise to the deviation of the dispersion curve from a straight line of slope $c$ and the retardation of the low frequency/low energy photons near the plasma frequency by kind of a friction the photons experienced in the homogeneous stationary density field. This is exactly the effect which the Higgs field plays in quantum field theory, applied here to a plasma in interaction with free space electromagnetic waves in classical electrodynamics.

In order to prevent a misconception, it may be important at this place to note that the photon mass as determined here, is \emph{not a universal rest mass} of the photons. There have been attempts in the past to find such a mass from observation of magnetic scintillations and plasma waves. These are highly questionable. \emph{Photons not in interaction} with some (Higgs) field \emph{remain massless.} 

The mass as determined above is the consequence of the interaction of the particular photons when passing across a plasma. Only as long as they interact, they carry it, and its precise value is determined by the Higgs field and the yukawa-coupling constant which, similar to field theory, determines the strength of the interaction. To be quite clear, this means that the photons of an electromagnetic wave which passes a plasma are indeed heavy, but they carry mass only as long as they interact with the plasma. The value of this mass depends on the density field $N_0$ and the yukawa-coupling constant, which may be different for different ways of interaction as for instance in the presence of a magnetic field as shown below. In contrast, in field theory like weak interaction the photons are permanently interacting with the Higgs field to become and remain heavy as long as they interact, and only during this period they are measured as the known heavy bosons.   In the absence of weak interaction they do not exist respectively remain massless. From that point of view, photonic mass is a temporary property which is maintained by the continuation of the interaction with the available Higgs field.

Identification of the Higgs field in plasma with the density field $N_0$ is rather satisfactory. In contrast to quantum field theory, however, it makes little sense to ask for a quartic potential of the kind $\Phi(N)-\Phi(0)= \alpha N^4-2\beta N^2$ of which the stationary Higgs field would be derived as its minimum state at $N_0=\sqrt{\beta/\alpha}$. In plasma the density is a real measurable quantity, which liberates us from any further inquiry.

\section{Plasma in external magnetic field}
Including an external quasi-stationary magnetic field complicates the calculation but leads to similar results. In this case a high-frequency electromagnetic wave in interaction with the plasma splits into two different polarisations, known as L-O and R-X modes in parallel or perpendicular propagation, respectively, with smooth transitions for increasing propagation angle. For our purposes restricting to parallel propagation, the R-L dispersion relation can be cast into the common form
\begin{equation}
\mathcal{N}^2=1-\frac{\omega_e^2}{\omega\:(\omega\mp\omega_{ce})}
\end{equation}
where the upper (lower) sign refers to R (L) modes, and $\omega_\mathit{ce}=eB/m_e$ is the electron cyclotron frequency in the magnetic field of strength $B$. this dispersion relation shows that for parallel propagation the two modes become cut off at long wavelengths $k\to0$ and finite cut-off frequencies 
\begin{equation}
\omega_\mathit{co}^\pm= {\textstyle\frac{1}{2}}\Big[(\omega_\mathit{ce}^2+\omega_e^2)^\frac{1}{2}\pm\omega_\mathit{ce}\Big]
\end{equation}
One may immediately conclude from this fact that the photons have attained an irreducible finite mass in the interaction with the plasma of the order of
\begin{equation}
M_\mathit{R,L}=\frac{\hbar\omega_\mathit{co}^\pm}{c^2}
\end{equation}
Moreover, the field the photons of the R-L modes interact with is now a combination of the homogeneous background density field $N_0$ and the external homogeneous magnetic field $B_0$, because these are the large macroscopic constant fields which enter the interaction of the massive plasma fields $\delta N, \delta\vec{A}$ with $N=N_0+\delta N, \vec{B}=\vec{B}_0+ \delta\vec{B}$ and $\vec{B}=\nabla\times\vec{A}, \vec{E}=-\partial_t\vec{A}, \vec{E}_0=0$. Since one may in addition assume that in the stationary thermodynamic equilibrium case the plasma will be in pressure balance between the thermal and magnetic pressure (a condition which might not necessarily be valid, however), the relation between $B_0$ and $N_0$ becomes $B_0=\sqrt{2\mu_0N_0T}$, where $T=$ const is the thermodynamic equilibrium temperature. It is not necessary here to again derive the wave equation for the massive plasma field. This is simple though more complicated by the fact that the external magnetic field is to be included. This equation is not necessary here; it suffices to identify the stationary (on the temporal and spatial scale of the electromagnetic wave $\delta\vec{A}$) the stationary classical Higgs field which leads to attribute the photon mass.

Manipulating on the above dispersion relation and interpreting it as an operator equation as done before, it can be cast into the approximate wave equation for the photon field
\begin{displaymath}
\bigg[\nabla^2-\bigg(c^2-\frac{\alpha}{2}V_{Ae}^2\bigg)^{-1}\frac{\partial^2}{\partial t^2}\bigg]\:\delta\vec{A}=\frac{\omega_e^2}{c^2}\bigg(1+\frac{\alpha}{2}\frac{\omega_\mathit{ce}}{\omega_e}\bigg)\:\delta\vec{A}
\end{displaymath}
where $\alpha=1,3$ for (R, L)-waves respectively, and $V_\mathit{Ae}=B_0/\sqrt{\mu_0m_eN_0}$ is the electron-Alfv\'en (or whistler) speed. Thus in this case the Higgs field is a more complicated combination of both, $N_0$ and $B_0$, and the Yukawa coupling constant is a variation of the former by the presence of the magnetic field.  Up to a small numerical difference which has its origin in the simplifications made in deriving the above wave equation, the photon mass obtained from this expression is the one inferred above from the dispersive cut-offs.  

The important point here is that the plasma provides the homogeneous large equilibrium value for the classical Higgs field which is responsible for the generation of a finite photon mass by retarding the photons in the interaction process and adding inertia to them. Hence, the Higgs mechanism is quite a general mechanism which is also present in classical systems as the interaction of photons with the reactive plasma in classical electrodynamics. Interestingly, this interaction also causes another effect, viz. the increase of the equilibrium dielectric constant of the plasma. This becomes clearer when investigating Alfv\'en waves in the following section. There the effect of the plasma is solely in the modification of the dielectric constant, while no photon mass is produced. This can be considered as the propagation of alfv\'enic massless fluctuations on the massive bulk plasma background field.

\section{No massive alfv\'enic photons}
One may be tempted to apply the same reasoning to Alfv\'en waves in an attempt to attribute an alfv\'enic photon mass to the photons which carry Alfv\'en waves. This turns out not to  be true, as can be easily shown. In spite of their confinement to the plasma, alfv\'enic photons remain massless.

The Alfv\'en wave is the non-resonant low-frequency branch of the electromagnetic ion-cyclotron mode. Kinetic plasma theory yields the dispersion relation of the latter 
\begin{equation}
\mathcal{N}^2\equiv\frac{k^2c^2}{\omega^2}=1-{\textstyle{\frac{1}{2}}}\frac{\omega_i^2}{\omega}\bigg(\frac{1}{\omega-\omega_{ci}}+\frac{1}{\omega+\omega_{ci}}\bigg)
\end{equation}
which, for $\omega\ll\omega_{ci}=eB_0/m_i$ the ion-cyclotron frequency in an ambient mean field $B_0$ and $\omega_i=e\sqrt{N_0/m_i\epsilon_0}$ the ion plasma frequency for an average quasi-neutral particle number density $N_0$, immediately yields if only the non-resonant part is considered
\begin{equation}\label{eq:disp-alf}
\mathcal{N}^2=1+\frac{\omega_i^2}{\omega_{ci}^2}, \quad \mathrm{with}\quad\frac{\omega_i^2}{\omega_{ci}^2}=\frac{c^2}{V_A^2}
\end{equation}
and $V_A=B_0/\sqrt{\mu_0m_iN_0}$ the Alfv\'en speed. This is the general (bulk) Alfv\'en dispersion relation 
\begin{equation}
\omega^2=\frac{k^2V_A^2}{1+V_A^2/c^2}
\end{equation}
where the usually neglected unity on the right hand side of the former equation has still been retained. It shows that the phase speed of the Alfv\'en wave is subluminal in particular for all physical group velocities $V_A<c$ for which the velocity ratio in the denominator can and is usually neglected. Of course, the Alfv\'en wave dispersion relation can be much easier derived from a simple one-dimensional fluid approach as was done first by \citet{alfven1942} in his seminal discovery paper.

However, the Alfv\'en wave in this frequency and wavenumber range $k<2\pi/ r_{ci}$, with $r_{ci}=v_i/\omega_{ci}$ the thermal ion gyroradius and $v_i=\sqrt{2T_i/m_i}$ the ion thermal speed at ion temperature $T_i$, is a genuinely electromagnetic wave, in this form not being subject to an electrostatic potential. Another condition is that the wavelength is shorter than the size $L$ of the plasma or curvature radius $R$ of the ambient magnetic field, which yields $kL,kR<1$. This fact indicates that its photons cannot have zero mass but, when interacting with the plasma, attain a finite mass, and it must also be this mass which allows them to enter the otherwise ideally conducting plasma deep into its bulk at any of its frequencies $\omega$. Any massless photons could penetrate the plasma at most up to the electron skin depth $\lambda_e=c/\omega_e$. This is a very interesting fact because it poses the question, in which way the massless photons of the electromagnetic wave have become loaded with mass until becoming a slow subluminal Alfv\'en wave which, because of its finite mass, propagates without any restriction inside the ideally conducting plasma. 

Usually this question is neglected and simply answered as that the plasma co-oscillates with the wave thus making the wave heavy. This answer is, however, not satisfactory because it does not explain how a co-oscillation of the plasma makes photons heavy, nor does it explain why photons can propagate at all in an ideal conductor which is by no means subject to any superconducting properties.

From the viewpoint of electrodynamics, Alfv\'en waves are simple electromagnetic waves. The free space equation applying to them is thus the wave equation
\begin{equation}
\Big(\nabla^2-\frac{1}{c^2}\frac{\partial^2}{\partial t^2}\Big)\:\delta\vec{A} =0
\end{equation}
where $\vec{A}$ is the vector potential which satisfies various conditions like $\nabla\times\vec{A}=\vec{B}$. However, comparing with the Alfv\'en dispersion relation and interpreting the frequency and wave number as operators, the wave equation for the Alfv\'en wave is cast into
\begin{equation}
\Big[\nabla^2-\frac{1}{c^2}\bigg(1+\frac{c^2}{V_A^2}\bigg)\frac{\partial^2}{\partial t^2}\Big]\:\delta\vec{A} = 0
\end{equation}
The effect of the plasma is thus to change the dielectric constant of the plasma in such a way that the velocity of light is reduced to become
\begin{equation}
c_A=\frac{cV_A}{\sqrt{c^2+V_A^2}} =\frac{c}{\epsilon_A}
\end{equation}
which for $c^2\gg V_A^2$ simply becomes $c_A\approx V_A$. It is interesting that no photon mass term is generated in this case of the internal bulk waves of the plasma. The plasma for the Alfv\'en wave behaves like a modified dielectric medium, not a medium on that the electromagnetic photons surf, interact with and gain a mass. Instead, the dielectric property of the plasma at these low frequencies far below plasma and cyclotron frequency, are an internal property of the medium like, for instance, the different dielectric property of water which causes light to become reflected and refracted. The difference is that the interior part of the plasma, i.e. the region below the resonance frequency of Alfv\'en waves, the ion cyclotron frequency, is not accessible to photons from the outside. Long before penetrating to such low frequency, the photons become cut-off in the R-L mode cut-offs and remain unrelated to Alfv\'en photons. 

This is very interesting and not easy to understand because it makes Alfv\'en photons not a  species different from ordinary massless photons, i.e. it does not make them heavy photons which have assumed a mass in the interaction with the plasma. Alfv\'enic photons remain ordinary massless photons even down to zero frequency and infinite wavelength (i.e. rather the extension of the plasma or the curvature radius of the magnetic field). For the relative alfv\'enic dielectric constant of the plasma this implies that it becomes
\begin{equation}
\epsilon_A\ =\ 1+c^2/V_A^2\ \approx\ c^2/V_A^2\,\gg 1
\end{equation}
This we already knew, and it is bound to hold only inside the plasma.


\newpage

\begin{acknowledgement}
This work was part of a Visiting Scientist Programme at the International Space Science Institute Bern. We acknowledge the hospitality of the ISSI directorate and staff.
\end{acknowledgement}



\end{document}